\documentclass[twocolumn,showpacs,preprintnumbers,amsmath,amssymb]{revtex4}

\usepackage[dvips]{graphicx}
\usepackage{bm}
\usepackage{epsfig}
\usepackage{dcolumn}

\begin{document}

\preprint{accepted at Phys.\ Rev.\ B}

\title{Role of semicore 
states in the electronic structure of group-III nitrides: An exact exchange study}
\author{A. Qteish and A. I. Al-Sharif}
\affiliation{Department of Physics, Yarmouk University, 21163-Irbid, Jordan}
\author{M. Fuchs and M. Scheffler}
\affiliation{Fritz-Haber-Institut der Max-Planck-Gesellschaft,
Faradayweg 4-6, D-14195 Berlin-Dahlem, Germany}
\author{S. Boeck and J. Neugebauer}
\affiliation{Department of Theoretical Physics, University of Paderborn, 
D-33095 Paderborn, Germany }

\date{\today}

\begin{abstract}

The bandstructure of the zinc-blende phase of AlN, GaN, InN is calculated employing 
the exact-exchange (EXX) Kohn-Sham density-functional theory and a pseudopotential 
plane-wave approach. The cation semicore $d$ electrons are treated both as valence
and as core states. The EXX bandgaps of AlN and GaN (obtained with the Ga 3$d$ 
electrons included as core states) are in excellent agreement with previous EXX 
results, GW calculations and experiment. Inclusion of the semicore $d$ electrons 
as valence states leads to a large reduction in the EXX bandgaps of GaN and InN. 
Contrary to common belief, the removal of the self-interaction, by the EXX approach, 
does not account for the large disagreement for the position of the semicore $d$ 
electrons between the LDA results and experiment. 

\end{abstract}

\maketitle

\section{Introduction}

The modern computational methods applied to condensed matter systems are
mainly based on Kohn-Sham \cite{KS} formalism of density-functional 
theory (KS-DFT). When used in conjunction with  both the local-density 
(LDA) and generalized gradient approximations (GGA) for the 
exchange-correlation (XC) potential, the KS-DFT approach provides a very 
efficient and successful tool to compute the total energy and other 
related ground state properties. However, calculated KS bandstructures 
of solids display several limitations of this approach. The most serious 
shortcoming is the so-called {\it bandgap problem}: the LDA bandgaps of 
semiconductors and insulators are between about 30 to more than 100\% 
smaller than the corresponding experimental values. Another problem 
is the underestimation of the binding energies of the semicore electrons. 
For the group-III nitrides and II-VI compounds, currently intensely 
studied materials due to their potential technological applications, the 
LDA results for the position of the $d$ bands are higher than experiment 
by 2 to 4 eV. The GGA electronic structures of these materials differ 
marginally from those of LDA. It should be noted that these deficiencies 
reflect on the one hand a fundamental issue: even the 
(unknown) exact single-particle KS eigenvalues will not directly correspond to 
excitation energies of the interacting electron system, except in special 
cases~\cite{wb83,com:slater}. As to the bandgap, it is well known \cite{Godby} 
that there is a discontinuity in the {\em {exact}} XC potential ($\Delta_{XC}$) 
when an electron is added to the system. In general the magnitude 
of $\Delta_{XC}$ is however unknown. The second, separate, issue is the effect of the
(inevitable) {\it approximation} for the XC functional. The LDA errors for bandgaps 
as well as for binding energies of semicore states have been attributed \cite{AQ} 
to a spurious self-interaction in LDA and GGA calculations \cite{PZ,com:slater}.

The best available technique for calculating the electronic structure of solids 
is the quasiparticle theory, within Hedin's $GW$ approximation \cite{Aulbur-rev}. 
The non-selfconsistent GW approach based on LDA results (LDA-GW) is found to 
reproduce very well the experimental bandgaps of standard $sp$ semiconductors 
\cite{Aulbur}. For group-III nitrides and II-VI compounds, the pseudopotential 
\cite{GWCdS,Rohlfing98} and all-electron \cite{Kotani02} LDA-GW methods are found 
to give good bandgaps (few tenths of an eV smaller than experiment), provided that 
the {\em {entire}} semicore shell is considered as valence in the pseudopotential 
approach. The positions of the $d$ bands are markedly improved within LDA-GW 
\cite{Rohlfing98,Kotani02}, although they remain systematically higher than 
experiment by about 1~eV.

A very interesting recent development is the application of the exact-exchange 
(EXX) approach \cite{Langreth83}, within the KS-DFT approach, to calculate 
the electronic structure of solids. In this approach the local -- KS -- exchange 
potential is treated exactly, i.e., the spurious 
self-interaction \cite{PZ} of the electron with its own charge density is 
eliminated. In the exchange-{\it only} KS-DFT formalism, the EXX potential is 
equivalent \cite{Sahni} to the variationally best local approximation (i.e.,
the optimized effective potential) to the nonlocal Hartree-Fock operator 
\cite{Sharp,Talman}. The EXX approach was first 
applied to crystalline solids by introducing a spherical shape approximation
of the potential by Kotani and co-worker \cite{Kotani94,Kotani95,Kotani96,Kotani98}. 
This approximate EXX approach is found to highly improve the calculated bandgaps 
of several semiconductors \cite{Kotani96} and the electronic structure of 
transition metals \cite{Kotani98}, with respect to the LDA results. More recently, 
a pseudopotential EXX approach has been introduced \cite{Stadele97,Stadele99}. 
The pseudopotential approach is free from the above shape approximation of the 
potential and the convergence with respect to the computational parameters 
can be easily controlled. The pseudopotential EXX approach is found to 
give \cite{Stadele99} bandgap energies in good agreement with experiment and 
GW results \cite{Aulbur,Fleszar}, for a large variety of semiconductors 
studied -- except for diamond. On the other hand, for noble-gas solids 
\cite{Fleszar04} the pseudopotential EXX approach gives bandgaps which are 
larger than those of LDA, but they are still much smaller than the 
experimental values. Thus, in addition to their own fundamental and practical 
importance, the EXX results provide a very good starting point for the 
non-selfconsistent GW (EXX-GW) calculations \cite{Patrick}.  

The purpose of this work is of threefold: (i) To perform pseudopotential EXX
calculations of the electronic bandstructure of the meta-stable \cite{ZBphase} 
zinc-blende (ZB) phase of GaN, AlN, InN.
In these calculations the Ga 3$d$ and In 4$d$ electrons are treated as
valence states. The electronic structure parameters of the ground state wurtzite 
(W) phase of the considered nitrides can be inferred from those of the corresponding
ZB structure \cite{Stampfl,Bechs}. (ii) To systematically analyze the effects of the 
semicore states 
on the bandstructure. To do that, an additional set of EXX bandstructures for GaN 
and InN are calculated, with the semicore states included in the core. (iii) To analyze 
and separate the changes that are due to the use of EXX pseudopotentials (PPs) from 
those due to the EXX treatment of the valence electrons. This is done by performing 
EXX bandstructure calculations employing LDA-PPs, and LDA calculations using both 
LDA- and EXX-PPs. 

It is worth noting that the bandgap of W-InN is still a matter of debate. The bandgap 
of the recently grown single crystalline hexagonal InN epilayers is found \cite{EgNew}
to be of 0.7-1.0\,eV, which is much smaller than the previous and widely accepted 
\cite{EgExp} experimental value  of 1.9\,eV. However, the smaller bandgap is far from 
being well established \cite{Shubina,Bechs2004}. A very recent experimental study suggested 
\cite{Shubina} that this small bandgap is a misinterpretation of the Mie resonances 
due to scattering or absorption of light in InN samples containing clusters of metallic 
In. Moreover, a tentative bandgap of 1.4\,eV has been given in Ref. \onlinecite{Shubina}. 
Therefore, accurate theoretical studies of the bandgap of InN are in order.

The paper is organized as follows. Sec. II contains a description of the 
computational method. Our results are reported and discussed in Sec. III. A summary 
of our main results and conclusions is given in Sec. IV.

\section{Computational Method}

To perform EXX band structure calculations, we implemented the pseudopotential 
EXX formalism \cite{Stadele99} in the SPHIngX code \cite{sphingx}, which employs 
plane wave basis sets and {\em ab initio} pseudopotentials. As noted above, EXX 
bandstructure calculations are performed for only the ZB phase of the considered 
nitrides. This is done at the experimental lattice constants (4.37, 4.50 and 
4.98 {\AA} for AlN, GaN and InN, respectively \cite{expLattConst}). 
We use the experimental rather than the theoretical lattice constants (which can 
be obtained by minimizing the EXX or LDA total energy) to allow for a direct comparison 
with the experimental bandgaps and to avoid shifts in the bandgaps due to 
deformation potentials. 

\vskip 0.5in

Table I: Pseudopotential parameters. The local part of the ionic pseudopotential is chosen 
to be one of its components, referred to by its orbital angular momentum ($l$). For the 
Ga$^{+3}$ and In$^{+3}$ LDA-PPs, reference energies for the Ga 4$d$ and In 5$d$ states of 
20 and 25 eV are used, respectively. 
\vskip 0.1in
\begin{tabular}{llllll}
\hline\hline
 Ion & approach & \multicolumn{3}{l}{core radius ($a_0$)}  & local   \\
     &          & $l=0$ & $l=1$ & $l=2$                    & part             \\ 
\hline
N$^{+5}$                   & EXX         &  1.5  &  1.5  &       & $l=1$   \\ 
N$^{+5}$                   & LDA         &  1.5  &  1.5  &       & $l=0$   \\ 
Al$^{+3}$                  & EXX \& LDA  &  1.9  &  1.9  & 1.90  & $l=2$   \\ 
Ga$^{+13}$ \& In$^{+13}$   & EXX         &  2.2  &  2.2  & 2.20  & $l=0$   \\ 
Ga$^{+13}$ \& In$^{+13}$   & LDA         &  2.2  &  2.2  & 2.20  & $l=1$   \\ 
Ga$^{+3}$                  & EXX         &  2.1  &  2.1  & 2.50  & $l=0$   \\ 
Ga$^{+3}$                  & LDA         &  2.1  &  2.1  & 2.20  & $l=0$   \\ 
In$^{+3}$                  & EXX         &  2.2  &  2.2  & 2.90  & $l=0$   \\ 
In$^{+3}$                  & LDA         &  2.2  &  2.2  & 2.35  & $l=0$   \\ 
\hline\hline
\end{tabular}

\vskip 0.5in

Norm-conserving scalar-relativistic EXX- and LDA-PPs of the involved elements
are generated by using the Troullier-Martins optimization method \cite{TM}. The 
EXX-PP's are constructed as described in Ref. \onlinecite{EXXPP}. Then, the generated 
pseudopotentials are transformed to the separable Kleinman-Bylander form \cite{KB}. 
The ground state electronic configuration is assumed for all the atoms considered.  Following 
Ref. \onlinecite{Stampfl}, only {\it s} and {\it p} components are included for N. 
The pseudopotential parameters used are listed in Table~I. 
The transferability of the pseudopotentials has been carefully tested, and they are 
found to be free from ghost states \cite{GhostStat}. The construction of the LDA-PPs and 
the testing of both the EXX- and LDA-PPs are performed by using the FHI98PP code \cite{Fuchs}. 

In the LDA calculations we used the Ceperley-Alder \cite{CA} exchange-correlation data as 
parameterized by Perdew and Zunger \cite{PZ}. In the present EXX calculations, the exchange 
energy and potential are treated exactly and the above LDA functional is used for those of
the correlation. Hereafter, unless otherwise specified, this combination (exact-exchange plus 
LDA correlation) will be referred to simply as EXX. Brillouin zone integration is performed 
on a regular $4\times4\times4$ Monkhorst-Pack \cite{MP} mesh. The KS wavefunctions are expended 
in terms of plane waves up to a cutoff energy of 60, 65 and 70 Ryd for AlN, InN, and GaN, 
respectively.  To calculate the 
independent particle polarizability, ${\chi}_{\rm o}$ (see Ref. \onlinecite{Stadele99}), 
plane waves up to an energy cutoff of 35, 45 and 55 Ryd are included for  AlN, InN, and GaN, 
respectively. The energy cutoffs used for the EXX calculations for GaN and InN, with the 
semicore $d$ electrons included in the core, are exactly as those of AlN. The k-point sampling 
and the energy cutoffs are carefully tested, and are found to give an excellent convergence. 
For example, reducing the ${\chi}_o$ energy cutoff of GaN from 55 to 45 Ryd, keeping the other 
cutoff fixed at 70 Ryd, changes the calculated eigenvalues by less than 0.01 eV. 
The calculation of the exact exchange potential involves also the unoccupied KS states, through 
$\chi_0$ and the Green's functions of the orbital shifts. To ensure convergence, we fully 
diagonalize the Hamiltonian in the plane wave representation and include all conduction 
states.

\section{Results and discussion}

In this section we report and discuss our EXX and LDA electronic structures of the ZB form 
of AlN, GaN and InN. We will also briefly address the modifications of their bandstructure  
when going from the ZB to W phases. The most prominent bandstructure parameters of the ZB 
phase of the three nitrides considered, obtained with the semicore $d$ electrons treated as 
part of the frozen core, are listed in Table~II. Whereas, those of GaN and InN, obtained with 
the semicore $d$ electrons treated as valence states, are listed in Table~III. The available 
experimental results for GaN are also shown in Table~III. As an example, we show in Fig. 1 
the LDA and EXX bandstructures of ZB-GaN. 

\subsection{Comparison between the electronic structures of ZB and W phases}

We focus here mainly on the differences in the bandgap and the position of the 
shallow $d$ bands ($E_d$) by going from the ZB to W phases, studied at the LDA 
level. For the W phase, we have also considered the experimental structural 
parameters \cite{expLattConst}. The obtained results can be summarized as follows.  
($i$) In agreement with previous LDA calculations \cite{Stampfl,Bechs}, 
the direct bandgap at the $\Gamma$-point ($E_g^{\Gamma}$) of the W phase of GaN 
and InN are larger than that of the corresponding ZB phase by about 0.2 eV. 
($ii$) The ZB-AlN is an 
indirect bandgap semiconductor with conduction bands minima at the X-points, in 
agreement with experiment and previous calculations \cite{Stadele99,Stampfl,Aulbur}. 
($iii$) The values of $E_g^{\Gamma}$ are almost identical in both ZB and W phases 
of AlN, in agreement with a recent LDA study \cite{Stampfl}. 
($iv$) The value of $E_d$ remains almost the same by going from the ZB to W phases, 
for both GaN and InN. For instance, the LDA results for $E_d$ of GaN (obtained using 
EXX-PPs), with respect to the valence bands maximum, are of -16.84 eV for the ZB~ phase 
and -16.87 eV for the W phase. 
In the following we will mainly concentrate on the 
EXX and LDA electronic structure of only the ZB phase of the considered nitrides.    
For this reason, the reference to the ZB phase will be suppressed, and that to the W 
form will be explicity stated.

\subsection{Bandgaps and the effects of the semicore $d$ electrons}

We will start our discussion with AlN where no semicore states are present.
Table II shows that the EXX results for the fundamental 
indirect bandgap ($E_g^X$) are in very good agreement with the experimental value
(5.11\,eV, see Ref. \onlinecite{Aulbur}), and with the GW results \cite{Aulbur}. 
For $E_g^{\Gamma}$, experimental 
results are only available for the W phase, and the most recent value is of 6.033\,eV 
(Ref. \onlinecite{EgAlN}). Since the values of $E_g^{\Gamma}$ are almost identical in 
the ZB and W structures of AlN, one can conclude that the EXX and GW 
results for $E_g^{\Gamma}$ are in good agreement with each other, and with the above 
experimental value.

We turn now to the electronic structure of GaN, obtained without the 
semicore $d$ states and using EXX-PPs (Fig.~1a). This figure shows that the energetic 
position and dispersion of the valence bands are weakly affected when going from the 
LDA to EXX. The most noticeable difference is in the upper valence bandwidth. 
The conduction bands, however, are significantly shifted upward in energy in the EXX 
calculations, with respect to the valence bands maximum. This upward shift increases 
the bandgap from 1.97\,eV (LDA calculation with EXX-PPs, see Table~II) to 3.52\,eV. 
This EXX bandgap is in excellent agreement with previous EXX calculations 
\cite{Stadele99,Aulbur}, LDA-GW calculations~\cite{Rohlfing98} (with Ga $3d$ electrons 
were included in the frozen core) and experimental data (3.2 -- 3.3 eV, see Table.~III).

To analyze the effects of the semicore electrons, we show in Fig.~1b the EXX and LDA 
bandstructures of GaN, obtained by treating these electrons as valence states. 
In this case a rather different behavior is found: Compared to the EXX calculation 
without semicore states, the bandgap closes by 0.64\,eV --- leading to a bandgap of 
2.88\,eV. This reduction can be easily understood as a consequence of the symmetry 
allowed anion $p$ - cation $d$ repulsion \cite{WZ88}, which pushes the upper valence 
bands states up in energy, leading to a smaller bandgap. This EXX bandgap is in remarkable 
agreement with the pseudopotential LDA-GW result (2.88\,eV) of Rohlfing 
{\it et al.} \cite{Rohlfing98}, obtained with the entire Ga semicore shell considered 
as valence. Furthermore, these pseudopotential EXX and LDA-GW results agree well with the 
mixed basis all-electron LDA-GW result (3.03\,eV) for W-GaN \cite{Kotani02}, after 
subtracting the difference in $E_g^{\Gamma}$ between the W and ZB phases of GaN 
($\sim{}0.2$\,eV, see above). This indicates that the inclusion of the entire semicore 
shell as valence is not necessary in the EXX and EXX-GW (see Ref. \onlinecite{Patrick}) 
calculations. This is largely due to the linearity of the EXX potential \cite{EXXPP} and 
the improved description of the semicore atomic $s$ and $p$ states of Ga by the EXX 
approach. The binding energies of these states are appreciably increased by the removal of 
self-interaction, as shown in Fig. 2. Compared to experiment, both EXX and LDA-GW approaches 
underestimate the bandgap of GaN (by about 0.4 eV). However, it has already been shown that 
a combination of these two methods (i.e., the EXX-GW approach) gives \cite{Patrick} bandgaps 
that are in very good agreement with experiment for GaN and some II-VI compounds. 

Concerning InN, the experimental bandgap (available only for W-InN) is still controversial,
as noted above. Our results listed in Tables~II and III for InN show a behavior which 
is qualitatively very similar to that of GaN: Inclusion of the semicore (In $4d$) states 
in the core gives an EXX value for $E_g^{\Gamma}$ of $\sim{}1.5$\,eV, while considering 
them as valence lowers significantly this bandgap to $\sim{}0.8$\,eV. Assuming a similar 
error as that of the bandgap of GaN (of about 0.4 eV) and taking into consideration
the difference in the bandgap of InN when going from ZB to W phases (see Sec. III.A),
a direct fundamental bandgap for W-InN of $\sim{}1.4$\,eV is estimated. This result 
is smaller than the previously accepted experimental value of $\sim{}1.9$\,eV, and 
larger than the very recent results (0.7-1.0\,eV, Ref. \onlinecite{EgNew}). 
Incidentally, it agrees well with the (tentative) experimental value of 1.4\,eV 
suggested by Shubina {\em {et al.}} \cite{Shubina}. However, it should be noted that 
the EXX-GW approach does not always give bandgaps that are larger than those of EXX
(Ref. \onlinecite{Aulbur}), and, hence, the present estimate should be also treated 
as tentative. 
 
It is interesting to note that the EXX bandgap of InN when including the semicore 
(In $4d$) states is significantly larger than in all-electron LDA-GW calculations
\cite{Kotani02,Usuda04}, which give a bandgap of $\sim{}0$\,eV, for the W phase. 
This is in contrast to the case of GaN where both approaches give almost identical 
results. The reason appears \cite{Usuda04} to be related to the fact that the 
dielectric function (which enters the GW) is calculated using LDA, which gives 
a semimetallic ground-state (negative bandgap, see Table~III) for InN. Thus, the 
calculated dielectric function is rather different 
from the real one which should be that of a semiconductor. It is well known that Ge 
has also a negative LDA bandgap (about -0.1 eV). However, since this bandgap is smaller 
than that of W-InN and the overlap between the conduction and valence bands is localized in 
a very small region around the $\Gamma$ point, this does not seem to affect the application 
of the LDA-GW method to Ge (see for example Ref. \onlinecite{Aulbur}). To address this issue 
in the case of InN two indirect LDA-GW approaches have been adopted in the literature: 
(i) Bechstedt and Furthm\"uller \cite{Bechs} have performed pseudopotential LDA-GW calculations 
with the semicore $d$ electrons included in the core.  The reduction in the bandgap due to 
the inclusion of these semicore electrons as valence is taken into account {\it {a posteriori}}: 
assuming the same value as found by LDA calculations. (ii) In the work of Usuda {\em {et al.}} 
\cite{Usuda04}, the bandgap has been opened by applying hydrostatic pressure, i.e., by 
reducing the InN 
lattice constant. The accuracy of these indirect LDA-GW approaches is difficult to assess. 
They lead to LDA-GW bandgaps of ZB-InN of 0.52\,eV (Ref. \onlinecite{Bechs}) and 
$\sim{}0.4$\,eV (Ref. \onlinecite{Usuda04}, obtained by subtracting 0.2
eV from the reported result for the W phase), which are smaller than the
EXX result, of 0.81 eV. 

\subsection{Energetic position of the $d$ bands} 

our LDA results for $E_d$ obtained using LDA-PP's, see Table~III, are significantly higher 
than the corresponding experimental data, in agreement with previous LDA calculations. 
This deficiency has been attributed to the incomplete cancellation of the self-interaction 
\cite{AQ}. It is therefore tempting to speculate that the EXX calculations, 
which are self-interaction free, would give improved $d$ band energies. A remarkable 
finding of the present study is that this is not the case: the EXX results are 
very close to those of LDA. In GaN the EXX result is only 1 eV lower than the
LDA results, while in InN it is about 0.1 eV higher. The LDA-GW approach, on the other
hand, is found to lead to a significant downward shift in energy of the $d$ 
bands in many II-VI and group-III nitrides \cite{Kotani02}. However, the GW results for
$E_d$ are systematically higher than experiment by about 1 eV. 

In order to understand the above unexpected EXX results for $E_d$ let us inspect 
the LDA and EXX eigenvalues of the isolated Ga, In, and N (pseudo)atoms, shown 
in Fig.~3. These results are obtained by using the same 
pseudopotentials and XC functionals as those used for the 
corresponding bulk calculations, see Sec. II. The important features to note from 
Fig.~3 are as follows. (i) All the eigenvalues shown are shifted downward in energy due 
to the removal of the unphysical self-interaction by the
EXX approach. (ii) The downward shifts in the eigenvalues of the N 2$s$ and 2$p$ 
states are comparable to those of Ga 3$d$ states, and larger than that of the In 4$d$ 
states. This explains why in the GaN and InN compounds, the EXX $d$ bands remain 
roughly speaking at the same energetic position with respect to the N $2s$ and 
$2p$ derived valence bands as in the LDA.
(iii) In the LDA calculations using EXX-PP's only the eigenvalues of the Ga 3$d$ 
and In 4$d$ states are appreciably shifted downward in energy, compared to the 
LDA calculations for all electrons (i.e., using LDA-PPs). It is interesting to 
note that these shifts (which are of 3.0 and 1.5 eV, respectively) closely  match 
the downward shift 
of the $d$ bands in the corresponding bulk calculations (see Table~III). This 
leads to values for $E_d$ in a rather good agreement with
experiment. Similar results are obtained for some II-VI compounds by Moukara 
{\em {et al.}} \cite{EXXPP}, who suggested that further improvements might be achieved if 
the EXX approach is applied to all electrons. The results of the present work show that 
this is not correct. 

\subsection{Valence band-widths}

We will focus mainly on (i) the width of the upper three valence bands, which mainly 
originate from the N 2$p$ states, referred to as "upper VBW". (ii) The width of the total 
valence bands (referred to as "total VBW"), defined as the energy difference between the 
top of the valence bands and the minimum of the lowest energy valence band which 
originates from the N 2$s$ states. This means that the cation semicore $d$ bands are not 
considered as part of the main valence bands.  
  
Fig. 1 and Tables~II and III show that the upper VBW decreases (by about 0.5 eV) by going 
from LDA to EXX calculations. The width of the lowest valence 
band (see Fig.~1a) also decreases, but by a much smaller amount. The reduction of the upper 
VBW  is consistent with previous EXX calculations \cite{Stadele99,Fleszar,Aulbur}. 
This feature can be explained \cite{Stadele99} as a direct consequence of the 
increase in strength of the effective potential seen by the valence electrons, due to 
the removal of the self-interaction by the EXX approach. The LDA-GW upper VBW's 
\cite{Aulbur,Rubio} are larger than those of EXX by about 1\,eV, for both AlN and GaN. 
For a discussion about the comparison between the EXX and LDA-GW results see the following
subsection.

Fig. 1 and Tables~II and III show that the total VBW is weakly affected by going from 
LDA to EXX calculations. The total VBW is affected through the reduction of the upper 
and lower VBW's and the difference in the downward shifts in energy of the involved 
bands, and these two contributions seem to cancel each other to a large extent. Table II 
show that, unlike the case of the bandgap, quasiparticle corrections have sizable 
effects on both the upper and total VBW's, in the case of both LDA (i.e., LDA versus 
LDA+GW results) and EXX (EXX versus EXX+GW results) approaches. The LDA-GW results 
\cite{Aulbur,Rubio} for the total VBW are appreciably larger than those of EXX. 
For GaN, both the EXX and GW results deviate significantly from the experimental 
value for the total VBW. The reason for such bad disagreement is unclear, 
but experimental uncertainties can't be ruled out.

\subsection{Further discussion}

In this subsection we discuss our EXX results in comparison with those of GW, 
from the point of view of the conceptual difference between the two 
approaches. Furthermore, we discuss the role of using consistent pseudoptentials
(i.e., EXX-PP's) in the EXX calculations, and comment on the discrepancies between 
the pseudopotential and all-electron EXX results.

We start with the comparison between the EXX and LDA-GW results. These two approaches 
provide different approximations to the XC self-energy operator. The LDA-GW 
approach is a many-body technique which leads to approximate quasi-particle or -hole 
energies, while the EXX method is a single-particle approach. As shown in this work 
and Refs. \onlinecite{Aulbur,Fleszar}) there is a very good agreement between the 
bandgaps calculated by both methods for wide range of semiconductors. This may 
suggest that for these systems the XC potential in the EXX approach (including LDA 
correlation) can be considered as a very good approximation to the LDA-GW self-energy 
operator, in the optimized effective potential sense \cite{Casida}. 
The conduction and upper valence-bands states are highly delocalized states and, thus, 
the effects of charge density relaxation by the removal (addition) of an electron from 
(to) these states are expected to be small. This is not the case for the more localized 
valence and semicore $d$ states, as the $\Delta SCF$ calculations show~\cite{wb83,ary96a}. 
This explains the good agreement between the EXX and LDA-GW results for the bandgaps, and 
the appreciable discrepancies between their results for $E_d$ and the valence band widths. 
Furthermore, this also implies that one should be careful when comparing the calculated 
EXX (or LDA) binding energies of the semciore $d$ electrons with the experimental photoemission 
data, since the latter include electronic relaxation effects. 
  
Now we address the question of why the EXX approach yields highly improved bandgaps, 
especially for the conventional $sp$ semiconductors. Following St\"adele {\em {et al.}} 
\cite{Stadele99}, we write the bandgap as 
\begin{equation}
E_g = E_g^{KS} + \Delta_{XC} = E_g^{EXX(X)} + \epsilon_{g,c}^{KS} + \Delta_{XC}.
\end{equation}
Here, $E_g^{EXX(X)}$ is the EXX bandgap obtained without any correlation, and 
$\epsilon_{g,c}^{KS}$ is the change in the bandgap due to the KS correlation potential. 
For conventional $sp$ semiconductors it has been shown that $E_g \approx E_g^{EXX(X)}$, 
and  $\epsilon_{g,c}^{KS}$ is quite small at the LDA or 
GGA levels ($\sim{}0.1$ eV). These results have led St\"adele {\em {et al.}} 
\cite{Stadele99} to conclude that $\Delta_{XC}$ is quite small, although the 
corresponding discontinuity in the EXX potential alone is very large, compared 
to the bandgaps of these systems \cite{Stadele99}. On the other hand, the 
Random-phase approximation (RPA) investigations \cite{GodbyRPA,Kotani98,Niquet} 
have shown that the KS bandgaps derived from these GW calculations are very close 
to those of LDA, for bulk Si and C. This means that RPA $\epsilon_{g,c}^{KS}$ is 
quite large. Thus, according to these RPA calculations, the EXX approach provide
improved bandgaps because of a large cancellation between $\Delta_{XC}$ and the 
errors in the LDA (or GGA) $\epsilon_{g,c}^{KS}$.   

The effects of employing consistent pseudopotentials in the EXX calculations (on the 
calculated bandgaps and valence band-widths) can be easily seen by 
comparing the EXX or LDA results obtained using both LDA- and EXX-PPs, listed in Tables II 
and III. The interesting features to note are: (i) if the same set of pseudopotentials (LDA 
or EXX) is used in both calculations, the bandstructure modifications by going from LDA to 
EXX calculations are almost independent on the employed pseudopotentials. (ii) For the three 
nitrides considered, these effects are quite small on the above properties, contrary to $E_d$, 
expect for the total VBW of InN obtained with the semicore $d$ electrons included as core 
states. This can be easily understood from the shifts in the eigenvalues of the pseudoatoms 
shown in Fig. 3. It should be noted that this is not always the case: the use of EXX-PPs in the 
LDA calculations leads to an increase in the bandgap of GaAs by about 0.6\,eV, and to a correct 
ordering of the conduction band minima of Ge \cite{Stadele99}. (iii) However, the advantages of 
using consistent pseudopotentials in the EXX calculations of the considered nitrides are clear: 
the comparison between EXX results obtained using EXX-PPs with experiment is better than those 
obtained employing LDA-PPs, especially for the the bandgap of GaN and $E_d$ of both GaN and InN.

After the completion of this work, all-electron full-potential EXX results have been reported 
\cite{Sharma}. These results are considerably different from those of pseudopotential EXX 
calculations, especially for the bandgaps of $sp$ semiconductors and the energetic position of 
the $d$ bands. For the latter the all-electron EXX results are in very good agreement with 
experiment, contrary to our results (see above and Ref. \onlinecite{Patrick}). These discrepancies 
have been attributed mainly to the core-valence interaction \cite{Sharma}. We strongly believe 
that such a conclusion is misleading because of two reasons: 
(i) We have shown (in Sec. III.B) that our pseudopotential EXX results for $E_d$ can be easily 
explained in terms of the eigenvalues of the corresponding isolated atoms, where core-valence 
interaction is not an issue. (ii) There are significant differences in the implementation of the
EXX scheme in the two approaches (see Refs. \onlinecite{Sharma,Stadele99} for details), which may 
account for the above discrepancies.  However, the reason behind these discrepancies can be 
clarified by performing pseudopotential EXX calculations with the entire semicore shell included 
as valence, which is currently under consideration.  

\section{Conclusions }

The exact-exchange (EXX) Kohn-Sham density-functional theory is used to calculate the 
electronic structure of the zinc-blende (ZB) form of AlN, GaN and InN, with and without 
including the cation semicore states. The effects of using consistent (i.e., EXX) 
pseudopotentials in the EXX calculations are also investigated. The changes in the 
bandgap and energetic 
position of the shallow $d$ bands by going from the ZB to wurtzite (W) forms have been 
studied, at the LDA level. Our EXX results for AlN and GaN, obtained with the semicore 
$d$ electrons treated as part of the core, are found to be in good agreement with similar 
previous EXX calculations, GW results and experiment. Treating the semicore $d$ electrons 
as valence leads to a large reduction in the calculated EXX bandgaps, and the results are 
in excellent agreement with the pseudopotential (when the entire semicore shell is 
included as valence) and all-electrons GW results. This may indicate that the inclusion 
of the entire semicore shell as valence is not necessary in the EXX and EXX+GW approaches.
Contrary to common belief, the removal of the self-interaction by the EXX 
approach, does not account for the large discrepancy for the position of the semicore $d$ 
bands between the LDA results and experiment. Most of the opening of the bandgaps of the 
considered nitrides in the EXX approach is due to its application to the valence electrons, 
and its application to the core electrons has pronounced effects on the position of the $d$ bands.  
The EXX method gives an appreciable bandgap for InN when the semicore $d$ electrons 
are included as valence, which provides the basis for more accurate GW calculations. 

\begin{center}
{\bf {Acknowledgments}}
\end{center}

we acknowledge stimulating discussions with Patrick Rinke and thank J. A. Majewski 
and P. Vogl for making their code for generating EXX pseudopotentials available to us. 
This work has been supported by the Volkswagen-Stiftung/Germany and the Deutsche 
Forschungsgemeinschaft (Research Group Nitride based quantum dot lasers).  


\newpage
\vskip 0.2in
Table II: Selected EXX and LDA electronic structure properties of the ZB phase
of the nitrides considered, obtained with the semicore $d$ electrons treated 
as core states. All tabulated quantities are in eV.
\vskip 0.1in
\begin{tabular}{lllll}
\hline\hline
         & LDA  & EXX   & LDA-GW      & EXX-GW     \\
\hline
\multicolumn{5}{l}{AlN} \\ 
\hline
 $E_g^{\Gamma}$&  4.27$^a$,       & 5.99$^a$     & 6.0$^{c,d}$   & 5.98$^c$   \\ 
               &  4.03$^b$        & 5.76$^b$     &               &                \\ 
               &  4.20$^{c,d,e}$  & 5.66$^{c,e}$ &               &            \\ 
 $E_g^X$       &  3.27$^a$        & 5.18$^a$     & 4.9$^{c,d}$   & 4.89$^c$   \\ 
               &  3.16$^b$        & 5.08$^b$     &               &                \\ 
               &  3.20$^{c,d}$    & 5.03$^{c,e}$ &               &            \\ 
               &  3.24$^e$        &              &               &            \\ 
upper VBW      &  5.91$^a$        & 5.27$^a$     &   6.7$^{c,d}$ &  6.29$^c$ \\
               &  5.96$^b$        & 5.33$^b$     &               &                \\ 
               &  6.0$^{c,d}$     & 5.24$^c$     &               &   \\ 
total VBW      & 14.81$^a$        & 14.59$^a$    & 17.00$^{c,d}$ & 15.32$^c$  \\ 
               & 15.16$^b$        & 14.95$^b$    &               &            \\ 
               & 15.10$^{c,d}$    & 14.85$^c$    &               &            \\ 
               & 14.74$^e$        & 14.86$^e$              &               &            \\ 
\hline
\multicolumn{5}{l}{GaN} \\ 
\hline
$E_g^{\Gamma}$ & 2.20$^a$        & 3.68$^a$      & 3.59$^f$     & 3.44$^c$  \\ 
               & 1.97$^b$        & 3.52$^b$      & 3.10$^{c,d}$ &           \\ 
               & 2.00$^c$        & 3.49$^{c,e}$  & 2.76$^g$     &           \\ 
               & 2.10$^d$        &               &              &           \\ 
               & 1.9$^e$         &               &              &           \\ 
               & 1.88$^g$        &               &              &           \\ 
upper VBW      & 6.85$^a$        & 6.30$^a$      & 6.88$^f$     &  7.17$^c$      \\ 
               & 6.82$^b$        & 6.23$^b$      & 7.6$^c$      &                \\ 
               & 6.8$^c$         & 6.23$^c$      & 7.8$^d$      &                \\ 
               & 7.4$^d$         &               &              &                \\ 
total VBW      &15.85$^a$        & 15.35$^a$     &  16.70$^c$   & 16.05$^c$      \\ 
               &15.82$^b$        & 15.63$^b$     &  17.80$^d$   &                \\ 
               &15.50$^{c}$      & 15.64$^{c,e}$ &              &                \\ 
               &16.30$^d$        &               &              &                \\ 
               &15.75$^e$        &               &              &                \\ 
\hline
\multicolumn{5}{l}{InN} \\ 
\hline
$E_g^{\Gamma}$ &  0.17$^a$       & 1.89$^a$      & 1.31$^h$     &         \\ 
               & -0.13$^b$       & 1.49$^b$      &              &                \\ 
upper VBW      &  5.52$^a$       &  4.88$^a$     &              &         \\ 
               &  5.67$^b$       &  5.09$^b$     &              &         \\ 
total VBW      & 13.98$^a$       & 13.82$^a$     &              &         \\ 
               & 14.50$^b$       & 14.37$^b$     &              &         \\ 
\hline\hline
\multicolumn{5}{l}{$^a$Present work, using LDA pseudopotentials.} \\ 
\multicolumn{5}{l}{$^b$Present work, using EXX pseudopotentials.} \\ 
\multicolumn{5}{l}{$^c$Ref. \cite{Aulbur}; $^d$Ref. \cite{Rubio}; $^e$Ref. \cite{Stadele99}; 
                  $^f$Ref. \cite{Rohlfing98};}    \\ 
\multicolumn{5}{l}{$^g$Ref. \cite{Palummo}; $^h$Ref. \cite{Bechs}. }   \\ 
\end{tabular}

\vskip 0.2in

\newpage 
Table III: Selected EXX and LDA electronic structure properties of the ZB phase
of GaN and InN, with the semicore $d$ electrons treated as valence states. For 
comparison some results for the W phase are shown, denoted by the superscript w. 
All tabulated quantities are in eV.
\vskip 0.1in
\begin{tabular}{lllll}
\hline\hline
          & LDA~~~~  & EXX~~~~  & LDA-GW~~~~      &   Expt.     \\
\hline
\multicolumn{5}{l}{GaN} \\ 
\hline
$E_g^{\Gamma}$ & 1.65$^a$      &   2.67$^a$  & 2.88$^c$      & 3.2$^d$   \\ 
               & 1.66$^b$      &   2.88$^b$  & 3.03$^{w,g}$  & 3.3$^e$   \\ 
     $E_d$     & -13.8$^{a}$   & -12.2$^a$   & -15.7$^c$     & -17.7$^h$ \\ 
               & -16.9$^b$     & -14.8$^b$   & -16.4$^{w,g}$ &           \\
               & -13.8$^{w,f}$ &             &               &           \\ 
upper VBW      & 7.32$^a$      &  7.03$^a$   &  7.33$^c$     &           \\ 
               & 7.03$^b$      &  6.63$^b$   &               &           \\
total VBW      & 16.00$^a$     & 16.04$^a$   &               & 14.2$^d$  \\ 
               & 16.04$^b$     & 16.00$^b$   &               &           \\
\hline
\multicolumn{5}{l}{InN} \\ 
\hline
    $E_g^{\Gamma}$  & -0.43$^a$     & 0.61$^a$   &  $\sim{}0.0^{w,f,g}$ &  0.7-1.0$^{w,j}$  \\ 
                    & -0.39$^b$     & 0.81$^b$   &                      &  1.9$^{w,k}$      \\
                    & -0.36$^i$     &            &               &                   \\ 
    $E_d$           & -13.5$^{a,i}$ & -12.2$^a$  &               &  -14.9$^{w,l}$    \\ 
                    & -14.9$^b$     & -13.4$^b$  &               &  -16.7$^{w,m}$    \\
                    & -13$^{w,f}$   &            &               &                   \\ 
upper VBW           &  6.10$^a$     & 5.82$^a$   &               &          \\ 
                    &  5.85$^b$     & 5.48$^b$   &               &           \\
total VBW           & 14.60$^a$     & 14.67$^a$  &               &          \\ 
                    & 14.71$^b$     & 14.73$^b$  &               &           \\
\hline\hline
\multicolumn{5}{l}{$^a$Present work, using LDA pseudopotentials.} \\ 
\multicolumn{5}{l}{$^b$Present work, using EXX pseudopotentials.} \\ 
\multicolumn{5}{l}{$^c$Ref. \cite{Rohlfing98}; $^d$Ref. \cite{EgGaN1}; $^e$Ref. \cite{EgGaN2}; 
$^f$Ref. \cite{Usuda04}; $^g$Ref. \cite{Kotani02}.}   \\ 
\multicolumn{5}{l}{$^h$Ref. \cite{Ding}; $^i$Ref. \cite{Bechs}; $^j$Ref. \cite{EgNew}; 
$^k$Ref. \cite{EgExp}; $^l$Ref. \cite{EdInN1}.}   \\ 
\multicolumn{5}{l}{$^m$Ref. \cite{EdInN2}.}   \\ 
\end{tabular}

\vskip 0.5in

\begin{figure}[p]
\includegraphics[width=6cm]{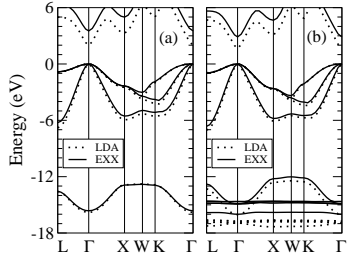}
\caption{\label{fig:GaN} Band structure of GaN calculated without (a) and with (b)
treating the Ga 3{\it d} electrons as valence states. The EXX pseudopotentials have
been used in both LDA and EXX calculations.} 
\end{figure}

\begin{figure}[p]
\includegraphics[width=6cm]{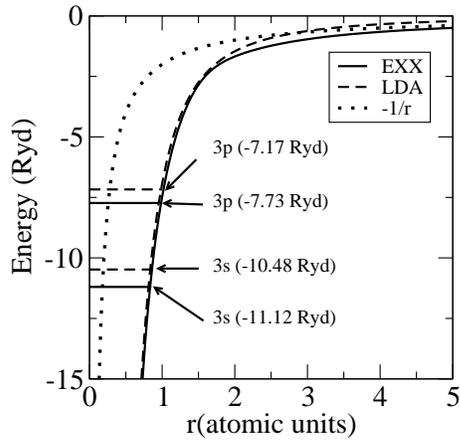}
\caption{\label{fig:pot} The atomic EXX and LDA all-electrons potentials of Ga. The right 
asymptotic $-1/r$ behavior of the EXX potential should be noted. The EXX and LDA eigenvalues
of the Ga 3$s$ and 3p are shown.} 
\end{figure}

\begin{figure}[p]
\includegraphics[width=6cm]{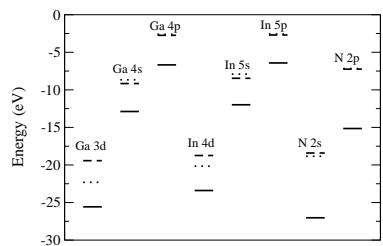}
\caption{\label{fig:atm-eng} Eigenvalues of the Ga, In and N (pseudo)atoms. Dashed lines: LDA 
calculations using LDA-PP's. Dotted lines: LDA calculations using EXX-PP's. Solid lines: EXX 
calculations using EXX-PP's.}
\end{figure}

\end{document}